\documentclass[twocolumn,epsfig]{revtex4}
\usepackage{epsfig,color}
\begin{document}
\title{Mass dependence of balance energy for different N/Z ratio}

\author{Aman D. Sood$^1$ }
\email{amandsood@gmail.com}
\address{
$^1$SUBATECH,
Laboratoire de Physique Subatomique et des
Technologies Associ\'ees \\University of Nantes - IN2P3/CNRS - Ecole des Mines
de Nantes 
4 rue Alfred Kastler, F-44072 Nantes, Cedex 03, France}
\date{\today}

\maketitle

\section*{Introduction}
 The investigation of
the system size effects in various phenomena of heavy-ion
collisions has attracted a lot of attention. The system size
dependences have been reported in various phenomena like
fusion-fission, particle production, multifragmentation,
collective flow (of nucleons/fragments) as well as its
disappearance, density, temperature and so on.
The collective
transverse in-plane flow has been investigated extensively
during the past three decades and has been found to depend
strongly on the combined mass of the system in
addition to the incident energy as well as
colliding geometry \cite{soodphd}. The energy dependence of
collective transverse in-plane flow has led us to its
disappearance. The energy at which flow disappears has been termed
as the balance energy (E$_{bal}$) or the energy of vanishing flow
(EVF). E$_{bal}$ has been found to depend strongly on
the combined mass of the system \cite{soodphd}.
 With the advent of radioactive ion beams the role of isospin degree of freedom on dynamics
 of heavy-ion collisions has been studied for the past decade. These studies are helpful to
 extract information about the asymmetric nuclear matter. In recent study Gautam et al. \cite{gautam2} has reported the different magnitude of isospin effects for different system masses. Here using IQMD model \cite{hart98} we aim to study the mass dependence of E$_{bal}$ for various N/Z ratios covering pure symmetric systems to highly neutron-rich ones. 

\begin{figure}[!t] \centering
\vskip 0.5cm
\includegraphics[angle=0,width=6cm]{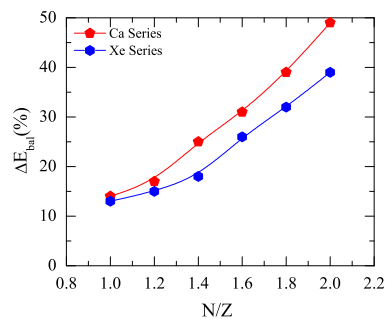}
\caption{\label{fig1} N/Z dependence of $\Delta$E$_{bal}$($\%$) for Ca and Xe series. Various symbols are explained in text. Lines are only to guide the eye.}
\end{figure}

\section*{Results and Discussion}
We simulate the reactions of Ca+Ca, Ni+Ni, Zr+Zr, Sn+Sn, and Xe+Xe
with N/Z varying from 1.0 to 2.0 in small steps of 0.2. The reactions are simulated at different incident energies around $E_{bal}$ in small steps of 10 MeV/nucleon. Collective transverse in-plane flow is then calculated at each incident energy. Then straight line interpolation is used to calculate the $E_{bal}$. Where $E_{bal}$ is the energy of vanishing flow. In Fig. 1, we display the percentage difference
($\Delta$E$_{bal}$($\%$)= $\frac{E_{bal}^{symm off}
-E_{bal}}{E_{bal}}$) between calculations for $E_{bal}$ without symmetry energy
and with symmetry energy as a function of N/Z for Ca (pentagons)
and Xe (hexagons) series. We see that the percentage difference
increases with increases in N/Z for both Ca and Xe series, which
shows that the effect of symmetry energy increases with increase
in N/Z. The increase is more sharp for Ca series as compared to
Xe, which indicates that with increase in N/Z the effect of
symmetry energy increases more sharply for Ca as compared to Xe
series. We also see that for N/Z = 1.0, the role of symmetry
energy is same throughout the mass range as far as E$_{bal}$ is concerned.\
\begin{figure}[!t] \centering
\vskip 0.5cm
\includegraphics[angle=0,width=6cm]{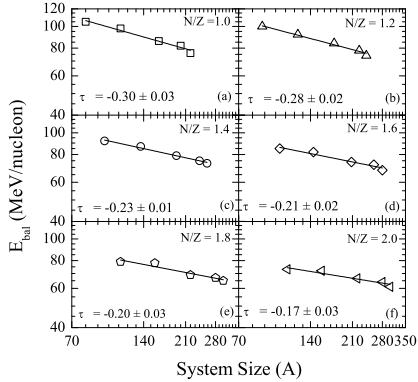}
\caption{\label{fig1} System size dependence of E$_{bal}$ for various N/Z ratios. Lines are of power law fit ($\propto$ A$^{\tau}$).}
\end{figure}
In Fig. 2, we display the system size dependence of E$_{bal}$
throughout the N/Z range from 1-2.We find that for each N/Z, E$_{bal}$
follows a power law behavior ($\propto$ A$^{\tau}$) with power law
parameter $\tau$ = -0.30$\pm$0.03, -0.28$\pm$0.02, -0.23$\pm$0.01,
-0.21$\pm$0.02, -0.20$\pm$0.03, and -0.17$\pm$0.03 for N/Z = 1.0 (open squares),
1.2 (triangles), 1.4 (circles), 1.6 (diamonds), 1.8 (open
pentagons), and 2.0 (left triangles), respectively. We find that
the value of $\tau$ decreases with increase in N/Z of the systems.
This is due to that fact that for higher N/Z ratio the effect of
symmetry energy is more in lighter masses (as discussed
previously), thus decreasing the E$_{bal}$ by larger magnitude on
inclusion symmetry energy in lighter masses which results in less
slope for higher N/Z ratio. Here we would like to stress that with increase
in N/Z for a given mass (compare mass of about 120 in Fig. 2(a) and 2(b)), the E$_{bal}$
decreases which is quite the opposite trend to the data \cite{pak98}. Since we have
taken the isotopes of a given element, where the data is for isobars and in Ref. \cite{gautam2} it has been shown clearly that Coulomb repulsion plays much more dominant role over symmetry energy in isospin effects (if one considers isobars) throughout the mass range and colliding geometry. Since Coulomb is repulsive so it lowers the $E_{bal}$ in isobaric system with small N/Z. For isotopes Coulomb will be same, the effect of symmetry energy (repulsive for neutrons and attractive for protons) will be more in isotopic system with larger N/Z thus lowering $E_{bal}$ for system having larger N/Z.
\section*{Acknowledgments}
This work has been supported by a grant from Indo-French Centre for the Promotion of Advanced Research (IFCPAR) under project no 4104-1.


\end{document}